\documentclass[prl,twocolumn,superscriptaddress,preprintnumbers,amsmath,amssymb]{revtex4-1}

\usepackage{graphicx}
\usepackage{dcolumn}
\usepackage{bm}
\usepackage{amsmath}
\usepackage{color}

\newcommand{\be}{\begin{equation}}
\newcommand{\ee}{\end{equation}}
\newcommand{\dd}{_{\rm D}}

\newcommand{\Eq}[1]{Eq.\,(\ref{#1})}
\newcommand{\eps}{\epsilon}

\begin{document}

\title{Time-dependent versus static quantum transport simulations beyond linear response}

\author{ChiYung Yam}
\affiliation{
Department of Chemistry, The University of Hong Kong, Pokfulam Road, Hong Kong}
\affiliation{ Bremen Center for Computational Materials Science, Am
  Fallturm 1a, 28359 Bremen, Germany}
\author{Xiao Zheng}
\author{GuanHua Chen}
\affiliation{
Department of Chemistry, The University of Hong Kong, Pokfulam Road,
Hong Kong}
\author{Yong Wang}
\author{Thomas Frauenheim}
\affiliation{ Bremen Center for Computational Materials Science, Am
  Fallturm 1a, 28359 Bremen, Germany}
\author{Thomas A. Niehaus}
\affiliation{University of Regensburg, 93040 Regensburg, Germany}
\email[Corresponding author: ]{ghc@everest.hku.hk, thomas.niehaus@physik.uni-r.de}
\date{\today}

\begin{abstract}

To explore whether the density-functional theory
non-equilibrium Green's function formalism (DFT-NEGF) provides a rigorous
framework for quantum transport, we carried out time-dependent density
functional theory (TDDFT) calculations of the transient current through two realistic
molecular devices, a carbon chain and a benzenediol molecule inbetween
two aluminum electrodes. The TDDFT simulations for the steady state
current exactly reproduce the results of fully self-consistent DFT-NEGF calculations  even beyond linear response.
In contrast, sizable differences are found with respect to an equilibrium, non-self-consistent treatment which are related here to differences in the Kohn-Sham and fully interacting susceptibility of the device region. Moreover, earlier analytical conjectures on the equivalence of static and time-dependent approaches in the low bias regime are confirmed with high numerical precision.

\end{abstract}

\maketitle

A first principles description of quantum transport at the atomic scale is usually
 based on the non-equilibrium Green's
function (NEGF) technique
\cite{Kadanoff1989}, which reduces to the traditional Landauer
transmission formalism \cite{Landauer1957,*Buttiker1986} for coherent transport. The device Green's function is
constructed in an {\em ad-hoc} manner from the Kohn-Sham (KS) single-particle Hamiltonian of ground state Density
Functional Theory (DFT) including the effects of the leads through suitable
self-energies
\cite{Cuevas2010}.
While such methods provide qualitative
results to understand the transport behavior, there are cases where the theoretical
steady state current differs from experimental
results by orders of magnitude \cite{Lindsay2007}. One reason for the discrepancy
is that the evaluated transmission function exhibits resonances at the
KS single particle energies, which are in general not physical.


Recently, several approaches based on {\em time-dependent} DFT (TDDFT)
have been put forward \cite{Tomfohr2001,*Baer2004,
  *Burke2005,*Kurth2005,*Aspuru2009,Zheng2007},
not only to treat genuine dynamical problems like AC
transport or transient effects, but also to remedy the problematic
situation in the steady-state case. For optical absorption spectra,
TDDFT excited state energies improve significantly on single-particle
energy differences that are obtained from the {\em static} ground state
KS equations. Since the screening properties of the device depend strongly on the optical
gap, one would expect also an improved description of the current.

\begin{figure}
\includegraphics[angle=0,scale=0.25]{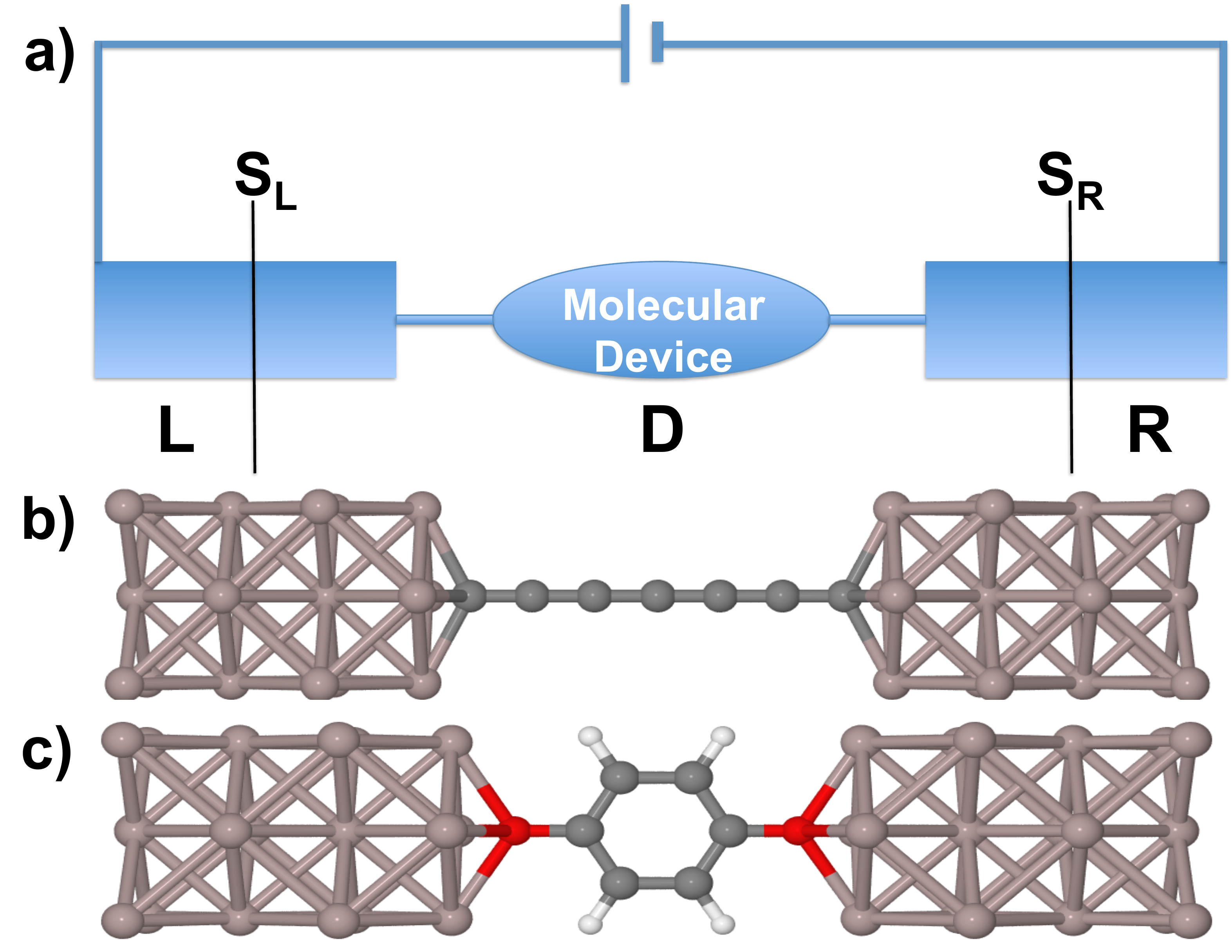}
\caption{\label{fig1} (a) Schematic device setup.
(b) Seven-membered carbon chain between Al nanowires in (001) direction.
Frontier carbon atoms in hollow position of the Al surface at 1.0 \AA\  distance.
Bond lengths (\AA): C-C = 1.32, Al-Al = 2.86.
(c) 1,4-benezenediol molecule inbetween Al nanowires.
}
\vspace{-0.5cm}
\end{figure}

In fact, dynamical corrections to the conventional DFT-NEGF picture have been predicted in the regime of
linear response \cite{Evers2004,*Sai2005,*Vignale2009}. Interestingly, these corrections were shown to vanish
for approximate xc functionals which are local both in time and
space. Such adiabatic functionals, like the adiabatic local density
approximation (ALDA a.k.a.\ TDLDA), are quite common in
many implementations of TDDFT for optical spectra due to their numerical simplicity. As shown by Stefanucci et. al. \cite{Stefanucci2007}
and Zheng et. al. \cite{Zheng2007}, TDLDA leads to the well-known
Landauer-B{\"u}ttiker formula for the steady state current if it can
finally be reached. It is thus argued that TDDFT and DFT-NEGF should result in the exact same steady
state current if the adiabatic approximation is adopted for the exchange-correlation
functional. We thus encounter  a paradox.

A related
open question is whether sizable differences between TDDFT and
DFT-NEGF may occur for local functionals {\em beyond} linear
response. The interest is here in the full I-V characteristics of the
device including the bias region around molecular resonances, which is
often used as fingerprint of the contacted molecule in
experiments. This regime is difficult to access with analytical
methods but amenable to numerical simulations.
Beyond linear response, more than one solution for steady state current is conceivable, and this
was observed for model systems \cite{Sanchez2006}. In such a case, the transient current may settle into different
steady currents depending on how the bias voltage is turned on. Also, it was argued that persistent
oscillating currents can exist and a steady state may never be reached. This is because
a device may have bound states with vanishing line widths and corresponding infinite life times.
The implications of such bound states for open system were investigated for a model system \cite{Khosravi2008}, and the
current through the model device was found to oscillate for as long as the simulation time.
To address these issues in this letter, we use a recently developed TDDFT method for open systems \cite{Zheng2007}
and compare results with those from DFT-NEGF calculations.

{\em Methodology.}---
For a two--terminal setup, the entire system consists of a device region ($D$),
and semi-infinite left ($L$) and right ($R$) electrodes as shown in
Fig~\ref{fig1}a. The
region $D$, where electron-scattering events take place, is thus an open
electronic system.

To obtain the reduced electronic dynamics of the device, we focus on the equation
of motion (EOM) of $\bm \sigma\dd$, the reduced single-electron density matrix for D,
\be
   i{\bm \dot\sigma\dd} = \left[ \bm h\dd, \bm \sigma\dd \right] -
   i\sum_{\alpha} \bm Q_\alpha(t). \label{eom-sigma-d}
\ee
where $\bm h\dd(t)$ denotes the KS Fock matrix of D, and the square bracket
on the right--hand side denotes a commutator.
$\bm Q_\alpha$ is the dissipation term due to the $\alpha$th electrode and
based on the Keldysh NEGF formalism \cite{Kadanoff1989}, it can be expressed as,
\begin{multline}
   \bm Q_\alpha(t) =
\\  - \int_{-\infty}^t d\tau \Big[ \bm G^r\dd(t,\tau)
   \bm \Sigma^<_\alpha(\tau, t) + \bm G^<\dd(t,\tau) \bm \Sigma^a_\alpha(\tau, t) \Big]
   + {\rm H.c.}
  \label{q-term-1}
\end{multline}
where $ \bm G^r_D$ and $ \bm G^<_D$ denote the time domain retarded and lesser
Green's function in the device region, and $ \bm \Sigma^a_\alpha$ and
$ \bm \Sigma^<_\alpha$ stand for the advanced and lesser self-energies of
lead $\alpha$.

The transient electric current through the interface $S_\alpha$ (the
cross section separating region $D$ from the $\alpha$th electrode)
can be evaluated via \cite{Zheng2007}:
\be
   I_\alpha(t) = - {\rm tr} \left[ \bm Q_\alpha(t)  \right].
\ee
%


If the bias potentials approach a constant value asymptotically
($V_\alpha^\infty \equiv V_\alpha(t\to \infty))$, a steady state may
develop. It has been shown that the steady state current can then be
recast in the form \cite{Stefanucci2007, Zheng2007}:
\begin{gather}
   I_L^\infty  = - I_R^\infty =\int d\eps \left[f_L(\eps - V_L^\infty) -
   f_R(\eps - V_R^\infty)\right]\, T(\eps)\nonumber \\
 T(\eps) =
 {\rm tr}\left[ \bm G\dd^r(\eps)
   \bm \Gamma_{R}(\eps)
   \bm G\dd^a(\eps)
   \bm \Gamma_{L}(\eps) \right], \label{t-of-e}
\end{gather}
 where $f_\alpha$ are Fermi distribution functions and
 $\bm \Gamma_{\alpha}(\eps) =  i \left[ \bm \Sigma^r_{\alpha}(\eps) -
   \bm \Sigma^a_{\alpha}(\eps)\right]$ describe the lead-molecule coupling.

 For non-local functionals, additional XC contributions
appear in the lead Fermi functions and give rise to an effective
bias which is smaller than the physical one
\cite{Evers2004,*Sai2005,*Vignale2009,
Stefanucci2007}. In contrast,
the current has Landauer form in the ALDA and seems to be equivalent to
the static DFT-NEGF result. However, the device Green's function
\begin{equation}
  \label{gree}
  \bm G\dd^r(\eps) = \left[ \eps \bm I  - \bm h\dd(\bm \sigma_D^\infty) -
    \bm\Sigma^r_{L}(\eps)- \bm\Sigma^r_{R}(\eps)\right]^{-1},
\end{equation}
is constructed from a Hamiltonian that depends on the steady-state
density. The latter is not necessarily the same as the density
obtained from a self-consistent DFT-NEGF treatment or even the
equilibrium one. To our knowledge, a general adiabatic theorem for open
systems is not available, though adiabaticity has been confirmed for
model potentials \cite{Scheitler1988}. Such a theorem would prove the equivalence of TDLDA and LDA-NEGF
approaches from \Eq{gree}. Hence, we provide a
numerical comparison of both approaches for some
realistic molecular devices in this letter.

As shown in Ref.~\cite{Zheng2007}, the expression for  $\bm Q_\alpha(t)$
in \Eq{q-term-1}
may be considerably simplified in the adiabatic wide--band limit (AWBL). It
depends finally only on the device density matrix $\bm \sigma\dd$, the
applied bias V(t) and the WBL parameters which characterize the lead
self-energies \footnote{Alternatively (but not employed in this study), a hierarchical
equation of motion approach \cite{Zheng2010} was recently developed, for which the
hierarchy terminates exactly at the second tier without any approximation.}. In this way the EOM is closed. At $t=0$, the
entire system is assumed to be in thermal equilibrium and the initial density matrix
$\bm \sigma\dd(t=0)$ is easily constructed using equilibrium Green's function (EGF)
techniques. Subsequently, bias potentials $V_\alpha(t)$ are applied to
the contacts and uniformly shift the lead energy levels.
The device density matrix $\bm \sigma\dd$ is then propagated according to
\Eq{eom-sigma-d}, where in each time step the Poisson equation is
solved to update the device Hamiltonian.

\begin{figure}
\hspace{-1cm}
\includegraphics[angle=0,scale=0.34]{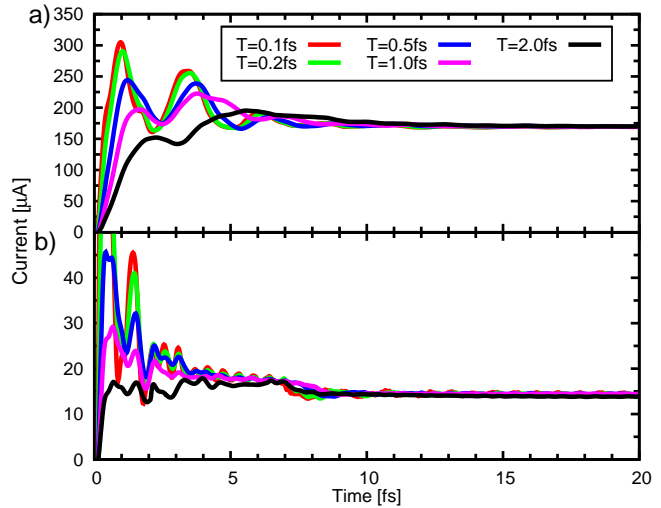}
\vspace{-0.5cm}
\caption{\label{time}
TDLDA transient current of a) carbon chain and b) benzenediol for
exponential turn-on of the bias voltage with different time constants $T$
and $V_0 = 3V$.}
\end{figure}

{\em Results.}---
We report the calculations of I-V curves for two molecular systems, a carbon
chain and a benzenediol molecule sandwiched between aluminum
leads in the (001) direction of bulk Al.  The structures are shown in
Fig.~\ref{fig1}(b) and  \ref{fig1}(c) respectively. Simulations for an Al wire were
also performed with similar findings as reported below. For each of the
molecular systems, we include explicitly 18 Al atoms of each contact in
the device region to ensure smooth potentials. The minimal
Gaussian basis set STO-3G is adopted in the calculations.
For the propagation, the fourth-order Runge-Kutta method is
employed with a time step of 0.02 fs to integrate Eq.~(\ref{eom-sigma-d}) in the time domain. The
ALDA is adopted for exchange and
correlation. The bias voltage is turned on exponentially at $t=0$ with a time
constant $T$, according to $V(t) = V_0(1 - e^{-t/T})$.

\vspace{-0.5cm}
\begin{figure}
\hspace{-1cm}
\includegraphics[angle=0,scale=0.33]{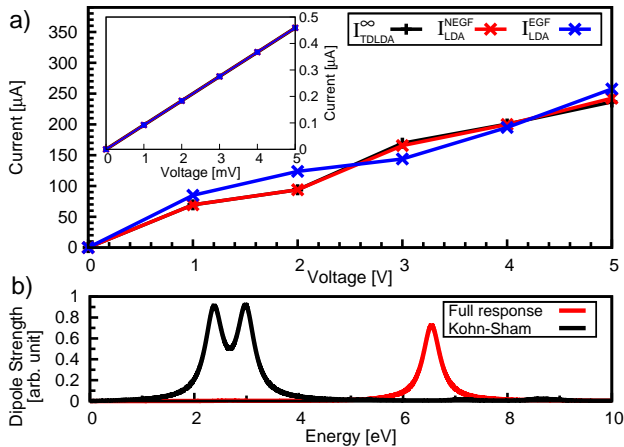}
\vspace{-1cm}
\caption{\label{alc7} (a) I-V curves of seven-membered carbon chain between Al nanowires
 in (001) direction at high bias and low bias (inset) regime.
(b) TDLDA optical spectrum of isolated and hydrogen capped carbon chain
(C$_7$H$_2$) (see text).}
\vspace{-0.5cm}
\end{figure}

\begin{figure}[hb]
\hspace{-1cm}
\includegraphics[angle=0,scale=0.33]{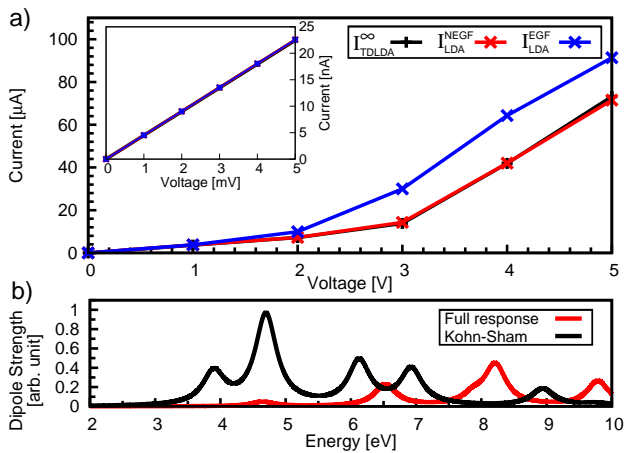}
\vspace{-1cm}
\caption{\label{alben} (a) I-V curves of 1,4-benzenediol inbetween Al leads at high
bias and low bias (inset) regime.
(b) TDLDA optical spectrum of isolated 1,4-benzenediol (C$_6$O$_2$H$_6$).}
\end{figure}

Fig.~\ref{time} depicts the transient currents for the two molecular systems at different values of
$T$ with $V_0 = 3$ V. Although the initial current traces differ, the final
steady-state current is the same in all cases. This was not obvious,
as the history of the applied bias voltage is included
in the formalism,
although the memory effect of the exchange-correlation functional is absent in the ALDA.
The establishment of the
steady state is ultrafast ($\simeq$ 10 fs) and occurs in every case. We
do not observe persistent current oscillations \cite{Khosravi2008}, as the
term $Q_\alpha$ in \Eq{q-term-1} ensures proper dissipation in the leads.

TDLDA simulations were then performed for several bias values in the
linear response, low voltage, regime. The asymptotic values of the
transient currents ($I^\infty_\text{TDLDA}$) were extracted in each case and
compared to static DFT calculations in the LDA using the same WBL
approximation for the leads in the insets of Fig.~\ref{alc7}(a) and Fig.~\ref{alben}(a). Two kinds of static DFT
results are reported. In the first case, the device Green's function
is determined fully self-consistently using the non-equilibrium density
($I^\text{NEGF}_\text{LDA}$). In the second case, the
equilibrium density is employed ($I^\text{EGF}_\text{LDA}$) and the Poisson equation is solved only once, which yields
a linear potential drop across the molecule.
For both studied systems all approaches yield identical results. This
provides numerical evidence for the earlier analytical arguments
\cite{Evers2004,*Sai2005,*Vignale2009} on the equivalence of static
and dynamic local DFT approaches in linear response.

In another set of simulations, the covered bias range was extended to
5 V. Finite bias simulations allow for the coverage of
the resonant tunneling regime in which the bias window contains one or
more of the molecular levels. Inspection of the transmission function (not
shown here) shows that transport
occurs predominantly through the unoccupied levels in both systems. For benzenediol,
the conducting molecular orbitals (MO) closest to the Fermi energy ($E_F$)
are located 0.92 eV above and 2.58 eV below $E_F$, respectively. The carbon chain exhibits
metallic behavior with a partially filled lowest unoccupied MO derived level \cite{Larade2001}.

The full I-V characteristics of both systems are summarized in
Fig.~\ref{alc7}(a) and  Fig.~\ref{alben}(a). The TDLDA
($I^\infty_\text{TDLDA}$)  and LDA-NEGF ($I^\text{NEGF}_\text{LDA}$)
current traces are virtually identical over
the whole bias range including the region of resonant
transmission. This level of agreement supports not only the validity
of the Landauer picture in the context of TDLDA (\Eq{t-of-e}),
but also shows that the TDLDA and LDA-NEGF
densities {\em are} in fact identical for realistic devices even for
strong bias voltage. Another reason to
study the high bias regime is the possible occurrence of multiple
solutions in the self-consistent LDA-NEGF treatment. As discussed by
S\'{a}nchez et al. \cite{Sanchez2006}, a dynamical approach would single out the most
stable steady-state solution in such a situation. Attempts to find
such  multiple  solutions were
unsuccessful for both devices. We also performed simulations with
increased molecule-lead distance in order to cover a larger parameter
space of coupling matrix elements. The transient currents show
oscillations with slower decay to the steady-state in these cases, but
again, the TDLDA results match the corresponding LDA-NEGF simulations exactly.

Comparing
now the TDLDA or LDA-NEGF results with LDA-EGF values and focussing first
on the bias region below 2 V, we find significant differences which
are more pronounced in the carbon chain. The LDA-EGF current is
initially higher, which can be traced back to a movement of the LUMO
resonance to higher energies in the self-consistent approaches. In
fact, TDDFT explicitly includes the potential change due to the
induced density, similar to the random-phase-approximation.  In the
context of harmonic perturbations in the optical range, this
leads to the well known shift of TDDFT excited states away from simple
KS single-particle energy differences. In order to quantify this
effect, we computed the optical spectrum of the isolated benzenediol
and carbon chain using a linear response TDDFT method \cite{Yam2003}.
Fig.~\ref{alc7}(b) and \ref{alben}(b) depict the results using the full response and a
non-self-consistent treatment employing the ground state
density, i.e.\ the KS response. Although in both systems the KS and TDDFT
spectra differ strongly, the deviations are higher in the carbon chain
with a shift of roughly 3 eV compared to a shift of 0.5 eV for
benzenediol. This finding is in line with the transport results for $V
< 2$ V. For larger values of the bias, the observed current traces are
more difficult to interpret since the transmission profile is
 altered in a non-linear fashion with respect to the equilibrium situation.

{\em Summary.}---
In this letter, we performed TDLDA transport simulations based on a time
propagation of the KS density matrix and compared these to
conventional static DFT results in the Landauer picture. The studied
systems comprise the ubiquitous benzenediol molecule with small equilibrium conductance
and an atomic chain with metallic conduction properties. In the regime
of linear response, the analytically predicted equivalence of
static and time-dependent DFT formalisms for local xc potentials could
be confirmed with high numerical precision. Beyond linear
response, TDLDA reproduces the result of LDA-NEGF calculations and deviates from static LDA only when
the equilibrium density is employed in the latter. This parallels the
energy shifts seen in going from the KS response to the full coupled
response in DFT absorption spectra. Although the transient current depends on
the specifics of applied bias voltage, a unique steady state is reached for
the same eventual bias voltage.

More advanced xc functionals beyond the ALDA are expected to lead to
an improvement in various ways. For example, non-local functionals provide
sizable dynamical corrections to the Landauer formula \cite{Sai2005}. Moreover,
functionals which exhibit a proper derivative discontinuity
refine not only the energetical position of molecular levels, but also the
molecule-lead coupling \cite{Toher2005, *Ke2007}. Despite of these short-comings, TDLDA is still
useful in the qualitative description of transport problems which
require a dynamical treatment, like the transient behavior of
molecular devices or photo-switches \cite{Yam2008}.

We thank the German Science Foundation (DFG, SPP 1243), Hong Kong Research
Grant Council (HKU700909P, 700808P, 701307P), Hong Kong University Grant
Council (AoE/P-04/08) and National Science Foundation of China (NSFC 20828003) for support.

\bibliography{Combined}

\end{document}